\documentclass[sigconf]{acmart}

\usepackage{booktabs}

\usepackage[normalem]{ulem} 
\usepackage{totcount} 
\newtotcounter{NonZSComments}

\newcommand{\todo}[1]{{\color{red} #1}}

\newcounter{TodoList}[subsection]
\newtotcounter{TodosLeft}

\newcommand{\DONE}[1]{}

\usepackage{float}
\usepackage{comment}
\usepackage{listings}
\usepackage{enumerate}
\usepackage{msc}
\usepackage{mathrsfs}
\usepackage{tikz, xstring, calc, pgfmath}
\usepackage{tcolorbox}
\tcbset{colback=gray!1!white}
\usepackage{mdframed}
\usepackage{mathtools}

\usepackage{setspace}

\usepackage{amsmath}
\usepackage{amsthm}
\usepackage{amsfonts}
\usepackage{amssymb}
\usepackage{bm} 

\newcounter{dummy} \numberwithin{dummy}{section}

\newtheorem{defn}[dummy]{Definition}
\newtheorem{lem}[dummy]{Lemma}

\usepackage{caption, subcaption}
\usepackage{dblfloatfix}

\newcommand{\mysubsect}[1]{\par\vspace{1.5ex}\noindent\textbf{#1. }}

\newcommand{\hash}[0]{\#}

\newcommand{\st}[0]{.\;}

\newcommand{\etal}{et~al.\ }

\newcommand{\factFont}[1]{\ensuremath{\mathsf{#1}}}

\newcommand{\tableSpacer}[0]{\quad - \;}
\newcommand{\alwaysSymb}[0]{\ensuremath{\checkmark}}
\newcommand{\verifSymb}[0]{\ \ensuremath{\checkmark^\ast}}
\newcommand{\violatedSymb}[0]{\ensuremath{\times}}

\DeclareMathOperator{\fst}{fst}
\DeclareMathOperator{\snd}{snd}
\DeclareMathOperator{\h}{h}
\DeclareMathOperator{\aenc}{aenc}
\DeclareMathOperator{\adec}{adec}
\DeclareMathOperator{\senc}{senc}
\DeclareMathOperator{\sdec}{sdec}
\DeclareMathOperator{\pk}{pk}
\DeclareMathOperator{\sign}{sign}
\DeclareMathOperator{\verify}{verify}
\DeclareMathOperator{\true}{true}

\newlength{\minwidth}

\newcommand{\mi}[1]{\ensuremath{\mathit{#1}}}
\newcommand{\type}[1]{\ensuremath \mathit{#1}}
\newcommand{\functionSet}[0]{\Sigma} 
\newcommand{\factSet}[0]{\mathsf{F}} 
\newcommand{\protocol}[0]{P} 
\newcommand{\ruleSet}[0]{R} 

\newcommand{\allTraces}[1]{\ensuremath \mi{Traces}(#1)} 

\newcommand{\logicOr}{\ensuremath \ \vee \ }

\newcommand{\ruleLabelFont}[1]{\ensuremath{\mathtt{#1}}}

\newcommand*{\xdash}[1][3em]{\rule[0.5ex]{#1}{0.55pt}} 

\newcommand{\makeRule}[3]
{
	\left[\  \begin{array}{@{}l@{}}#1\end{array}\ \right ] 
	\hspace{-1ex}%
	\xrightarrow{\substack{#2}}%
	\hspace{-1ex}%
	\left[\  \begin{array}{@{}l@{}}#3\end{array}\ \right] 
}

\newcommand{\makeNamedRule}[4]
{
    \ruleLabelFont{#1} :=
	\left[\  \begin{array}{@{}l@{}}#2\end{array}\ \right ] 
	\hspace{-1ex}%
	\xrightarrow{\substack{#3}}
	\hspace{-1ex}%
	\left[\  \begin{array}{@{}l@{}}#4\end{array}\ \right] 
}

\newcommand{\tls}{TLS Middlebox Extensions}

\newcommand{\ltk}{\mi{ltk}}

\usepackage{hyperref}
\hypersetup{colorlinks=true, citecolor=blue,linkcolor=blue}

\input{multisetrules}

\newif \ifshowauthor
\showauthortrue

\newif \ifauthorcomment
\authorcommenttrue

\newif \ifshowpagenumber
\showpagenumbertrue 

\setcopyright{none} 
\renewcommand\footnotetextcopyrightpermission[1]{} 

\acmConference{ArXiV}{}{}
\acmYear{2022}
\begin{document}

\fancyhf{} 
\fancyfoot[C]{\thepage}

\title{Modelling Agent-Skipping Attacks in Message Forwarding Protocols}

\author{Zach Smith}
\email{zach@almou.se}
\author{Sjouke Mauw}
\email{sjouke.mauw@uni.lu}
\affiliation{%
  \institution{University of Luxembourg}
}
\author{Hugo Jonker}
\email{hugo.jonker@ou.nl}
\affiliation{%
	\institution{Open University of the Netherlands}
}
\author{Hyunwoo Lee}
\email{lee3816@purdue.edu}
\affiliation{%
	\institution{Purdue University}
}

\begin{abstract}
Message forwarding protocols are protocols in which a chain of agents
handles transmission of a message. Each agent forwards the received
message to the next agent in the chain. For example, TLS middleboxes act
as intermediary agents in TLS, adding functionality such as filtering or
compressing data. 

In such protocols, an
attacker may attempt to bypass one or more intermediary agents. Such an
\emph{agent-skipping attack} can the violate security requirements of the protocol.


Using the multiset rewriting model in the symbolic setting, we construct
a comprehensive framework of such path protocols. In particular, we
introduce a set of security goals related to \emph{path integrity}: the
notion that a message faithfully travels through participants in the
order intended by the initiating agent. We perform a security analysis
of several such protocols, highlighting key attacks on modern protocols.
\end{abstract}


\maketitle

\section{Introduction}
\label{sec:introduction}

The complexity of communication systems often necessitates the use
of simplifying assumptions in order to enable functional models.
For example, we typically treat internet communication as two-party
protocols, focusing on the end-to-end security requirements. However,
modern protocols often involve the addition of intermediate agents in
order to enhance functionality. In these settings, messages are
forwarded down a \textit{path} of connected agents.
In secure routing protocols~\cite{icingRouting, optRouting},
agents might simply forward incoming messages to the next agent after
inspecting the packet headers. However,
some advanced protocols require agents to actively participate in some form.
For example, the TLS protocol~\cite{tls12,tls13} is used for an
overwhelming majority of modern web communication. TLS users often
use services that intercept messages, redirecting or modifying their
contents, such as load balancers and firewalls. Such \emph{TLS middleboxes} 
necessitate modifications to
the TLS protocol in order to support multiple parties. Currently,
the most common approach in handling this is known as Split
TLS~\cite{splitTLS}, in which the TLS session is ``split'' into a
series of completely disjoint sessions between each pair of intermediate
agents. Security concerns about Split TLS~\cite{killed,impact,intercept} have
led to new solutions, such as mcTLS~\cite{mctls}.

Mixnets and Onion Routing protocols also involve sending messages along
a path. Such protocols originate from Chaum~\cite{chaummixnet}. Modern
protocols such as The Onion
Router~\cite{tor} and Sphinx~\cite{sphinx} are built in a similar way.
These protocols create layered messages (``onions''), where each layer
contains information specific to one agent in the chain. This agent then
peels off that layer and forwards the remaining onion to the next agent.
A novel application of this is in the field of payment networks, e.g.,
the Lightning Network~\cite{lightning}. In this system,
off-chain payment channels between pairs of agents replace
traditional transactions. This leads to the concept of chained payments, in which
an agent can send funds to a peer through a connected series of such channels,
where each intermediate agent collects a small fee.

\begin{table*}[tb]\centering
\caption{Protocols which rely on Path Integrity for security}
\begin{tabular}{lp{4.1cm}p{5.3cm}l}
    \toprule
    \textbf{Protocol Family} & \textbf{Example Protocols} &  \textbf{Relevance of Path Integrity} & \textbf{Example Attack} \\
    \midrule
        Secure Routing &
        OPT~\cite{optRouting}, ICING~\cite{icingRouting}, RPL~\cite{RPLRouting}
        &
        Poisoning network topology &
        Tunelling~\cite{routerTunelling}
        \\
        Mixnet &
        Chaum~\cite{chaummixnet}, TOR~\cite{tor}, HORNET~\cite{hornet}  &
        Privacy of path compromised &
        RAPTOR~\cite{raptorTorAttack} 
    \\
        Payment Network &
        Lightning~\cite{lightning}&
        Transaction fee skimming & 
        Wormhole~\cite{lightningWormhole}
    \\
    Middlebox-Enabled TLS &
        mcTLS~\cite{mctls}, mbTLS~\cite{mbtls}, maTLS~\cite{matls}, meTLS~\cite{metls} &
        Bypass functionality-enhancing services &
 
        \\
        
    \bottomrule
    
\end{tabular}
\label{fig:protocolRelevanceTable}
\end{table*}

\begin{figure}[h]
    \centering
    \includegraphics[width=\columnwidth]{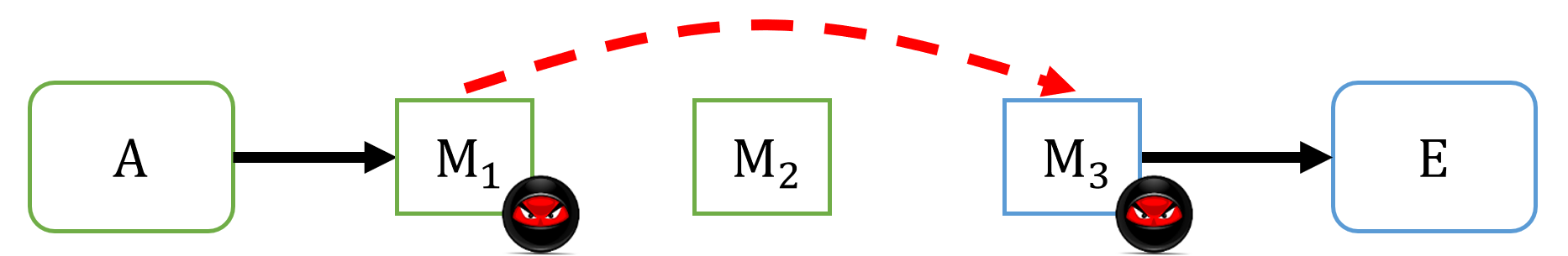}
    \vspace{-5ex}
    \caption{An agent-skipping attack: agents $M_1$ and
    $M_3$ collude to bypass $M_2$ using an out-of-band channel.}
    \label{fig:basic-attack-intuition}
\end{figure}

Several \emph{agent-skipping attacks} have been found on such
protocols~\cite{sensorWormhole,lightningWormhole}. Intuitively, these
attacks arise from the use of shortcuts: redirecting or modifying messages
in order to bypass one or more agents in the chain. The basic structure of
an attack is given in Figure~\ref{fig:basic-attack-intuition}.

The impact of a skipping attack depends on the specific setting. For TLS,
the principal agents want to ensure that their middleboxes are being
respected. For example, it must not be possible to bypass a content filter,
lest malicious injected code could reach an endpoint. 
For onion-style
protocols, the sender wishes to preserve privacy by guaranteeing that the
message travels only between the trusted intermediaries. Finally, 
for payment networks, intermediate agents must be assured that if they assist
in a payment by forwarding a message, they are guaranteed to receive their
transaction fees during the resolution stage.
In order to protect against such attacks, protocols require a notion of
\textit{path integrity}. Participating parties must be sure that 
all messages follow the intended path. This paper is focused on
building a simple and modular
framework for verifying such
a property.


\mysubsect{Contributions}
Our contributions are as follows:
\begin{itemize}
    \item We highlight the threat posed by message skipping
    attacks on a wide domain of protocols, including an in-depth case study
    on the mbTLS extension to TLS.
    \item We introduce a symbolic framework, built
    upon the multiset rewriting model, that can
    be used to describe the structure of
    `path-based' protocols.
    \item We give a formal definition of the
\textit{path integrity} security goal inside
this framework.
    \item We provide a collection of models
    of a range of protocols from the literature
    using the Tamarin prover tool, showing the applicability
    of our framework.
\end{itemize}


\section{Background and Related Work}
\label{sec:relatedWork}


We begin by briefly highlighting examples of failures
of Path Integrity across the literature, in Section~\ref{subsec:path-integrity-attacks}.
In Section~\ref{subsec:general-related}, we discuss existing formalisms
of related security goals.
Finally, Section~\ref{sec:tls-middleboxes} contains a case study of
the middlebox-enabled TLS extension mbTLS~\cite{mbtls}, to highlight
a specific scenario where Path
Integrity is relevant, and provide
an intuitive definition.

\subsection{Attacks on Path Integrity}
\label{subsec:path-integrity-attacks}

Table~\ref{fig:protocolRelevanceTable} contains a list of protocol
families of interest, as well as the consequences of skipping attacks.

\mysubsect{Secure Routing}
Secure Routing protocols form the simplest class of protocols where a notion
of Path Integrity is a goal -- with several different well-studied and 
comparable definitions,
including Path Enforcement~\cite{secureRoutingSurvey} or Path Compliance.
In these protocols, a path is often set by using a set of
path headers, along with some series of verification checks at intermediate
nodes. In many cases, a key focus is the process of path selection
and building an accurate view of the network topology.
Recently, several attacks have been identified on such protocols
for IoT devices~\cite{rplAttack},
which have constrained computational capacity~\cite{RPLRouting}.

\mysubsect{Mixnets}
Mixnets are a variant of Secure Routing protocols which sacrifice the ability
of individual nodes to validate the path in favour of enhanced privacy features.
Each node is made aware only of the previous and next node in the path, often
using a multiply-encrypted bundle containing the relevant headers. In this case,
an attacker injecting additional nodes in the path can violate privacy goals, as
they are able to infer the identity of the other intermediate agents.
There are many network-based attacks on Mixnet protocols, for example, using traffic analysis. In the case where an adversary is granted additional
privileges (for example, identity spoofing), interception attacks may exist~\cite{raptorTorAttack}.

\mysubsect{Payment Networks}
Payment Networks are Layer-2 cryptocurrency protocols for facilitating faster payments. The most notable is the Lightning Network~\cite{lightning}
for Bitcoin. Agents set up long-term payment ``contracts'' using Bitcoin, and
payments take place by re-negotiating this existing contract, allowing them to take place off-chain.

In order to send funds to a partner with whom you do not have an active contract,
the Lightning Network allows chained payments: agents may publicly declare the
presence of an open contract in order to allow others to route payments through it.
Informally, if $A$ wishes to pay $C$ an amount $n$, he makes a deal with $B$ such
that $A$ will pay $B$ $n$ if and only if $B$ pays this same amount to $C$. $B$ can then charge some small transaction fee for performing this service.

The atomicity of this transaction is protected by the use of a hashed challenge $h(x)$
(and associated preimage $x$). An agent can only retrieve their share of the funds
if they can produce the preimage to the next agent in the path. Although the Lightning protocol is built on top of the Sphinx~\cite{sphinx} Mixnet,
the shared value $h(x)$ removes the privacy benefits -- and worse, revealing the preimage $x$ causes a Wormhole attack, as shown by
Malavolta \etal~\cite{lightningWormhole}.
This allows colluding parties to skim
the transaction fees of agents between them in the path.


\mysubsect{Middlebox-Enabled TLS}
Unlike the other sets of protocols, a key focus in Middlebox-Enabled TLS protocols
is that intermediate agents generally have full access to the message payload.
As such, a key component of their design is in giving the endpoints control over
who has access to the necessary encryption keys, to avoid the leaking of sensitive
data.
We will discuss these protocols in more detail in Section~\ref{sec:tls-middleboxes}.

\subsection{Existing Formalisations}
\label{subsec:general-related}


Although the notion of Path Integrity is relevant to several domains, the
majority of the literature around this topic is restricted to Secure Routing
protocols, where it is the primary security goal.

Several formal treatments have been made for modelling such
protocols~\cite{pathModel}. This includes the work of
Zhang \etal~\cite{coqIntegrity}, who verify the OPT protocol~\cite{optRouting}
using an embedding of the LS$^2$ logic into the Coq~\cite{coq} tool. More
recently, Klenze \etal~\cite{basinRouting} built a more general approach using
the theorem prover Isabelle/HOL. They use this approach to verify several
secure routing protocols including ICING~\cite{icingRouting}.
%
%
In this work we highlight the fact that this security goal is in fact relevant
to several other protocol domains. Instead of using a general-purpose theorem
prover, we design a framework that is directly compatible with the
security tool Tamarin~\cite{tamarin}. This framework allows us to 
specify protocols from all of the above domains, using the standard Dolev-Yao~\cite{dolevyao}.
adversary in a symbolic setting, rather than the computational settings
previously used.


For this work we assume that the intended message path is fixed during the
setup of the protocol -- i.e. the initiating agent has selected the full path.
We do not make any assumptions about the knowledge of the intermediate agents
Though the path selection process often presents unique
problems~\cite{ariadneRouting,lightningRouteSelection,splitPayments},
we consider it out of scope of this work, as it is often domain-specific.

\subsection{\tls}
\label{sec:tls-middleboxes}

\begin{figure}[t]
    \centering 
    \includegraphics[width=\columnwidth]{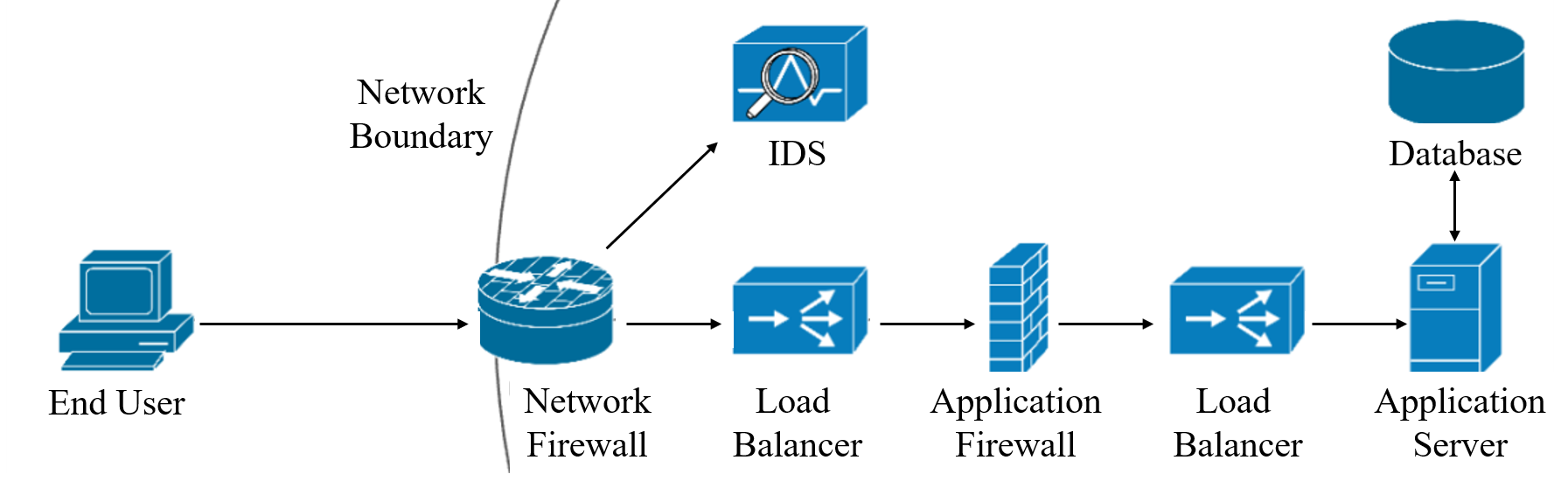}
    \caption{Example Network Architecture. Several intermediate applications
    operate between the Network Boundary and the Application Server.
    }
    \label{fig:mbtls-network-structure}
\end{figure}

Transport Layer Security (TLS)~\cite{tls12,tls13} is the de-facto standard for
end-to-end security on the web.
In some cases, endpoints may introduce 
TLS \textit{middleboxes} -- intermediate agents
that attempt to add functionality to a TLS session, 
such as by filtering undesirable traffic or compressing data. 
In order to perform these functions, such middleboxes require full access to the (encrypted) message
payloads, thus requiring that they are active participants in the
TLS session.

\mysubsect{Example Network}
In order to motivate the discussion in this section, consider
the example network infrastructure given in
Figure~\ref{fig:mbtls-network-structure}. Requests to a
web-facing service are passed through a series of middleboxes
before reaching the application servers. 
Intermediaries include Load Balancers which control the flow
of messages to multiple servers in parallel, and Application
Firewalls.

Traditionally, enabling these middleboxes is achieved using a
process known as SplitTLS~\cite{splitTLS}, which requires endpoints
to grant any middleboxes access to their certificate private keys (for a server),
or to install a root certificate (for a client).
This allows the middleboxes to impersonate the protected
application server.

However, SplitTLS has been shown to degrade the
security of a TLS session.
This could be caused by TLS middleboxes that are either incorrectly 
or maliciously implemented~\cite{impact, killed, modification, lenovo}.
For example, a middlebox may support insecure or deprecated ciphersuites,
allowing an adversary to perform a man-in-the-middle attack to degrade
the security of the entire session.


As an alternative to SplitTLS, a series of middlebox-enabled
TLS schemes have been proposed~\cite{matls, mbtls, mctls, metls}. These avoid the problems associated with
sharing keys by instead increasing the visibility of these
middleboxes and allowing the end user to confirm the authenticity of the application server directly.

\mysubsect{Security Goals of \tls}
Middlebox-enabled TLS extensions aim to achieve three main goals:

\begin{itemize}
    \item Allow middleboxes to identify themselves as active
    parts of a TLS session, rather than creating split sessions.
    \item As a result, avoid dangerous sharing of critical
    private keys (and their associated certificates) between
    multiple devices.
    \item Monitor and regulate the actions that middleboxes take.
\end{itemize}

These goals can be achieved in several ways.
Generally, approaches involve some combination of the following
strategies:

\begin{itemize}
    \item Establishing a secure channel between the two endpoints
    as part of the initial handshake, in order to exchange session-critical data.
    \item Delegation of session keys from endpoints to middleboxes,
    rather than piecewise establishment between pairs of agents.
    \item Addition of a ``modification log'' to messages, showing
    which middleboxes have modified a message in-transit.
\end{itemize}

%

In this work, although we allow for dishonest middleboxes
who release their keys, we assume that all honest agents
are willing only to run the specific TLS variant in
question, and follow it faithfully.
Further, our analysis considers only the record phase of TLS, where application data is secured with keys established during the
handshake phase of TLS. We assume the handshake was correctly performed.
Indeed, our security claims follow from the order of agents that is believed
to have been established during this handshake.



Although significant security analysis has been put into
the core TLS protocol suite~\cite{tls13analysis1, tls13analysis2}, the security
discussion of middlebox-enabled extensions is still
somewhat limited. 
Some formal models have been created for accountable proxying,
such as those of Bhargavan \etal\cite{3acce, acce-ap}. However,
these focus more on end-to-end authentication
guarantees in the presence of proxies, avoiding discussion
of path integrity. They use a computational setting, rather
than a symbolic model such as the one we present here.
One middlebox-enabled TLS extension, 
maTLS~\cite{matls},
provides a definition of path integrity for their specific
use-case, while we aim to produce a broad-scope definition.

\begin{figure}[b]
    \centering
    \input{diagrams/mbtls-setup-attack.tex}
    \caption{Overview of the mbTLS protocol}
    \label{fig:mbtls-skipping-attack}
\end{figure}

\mysubsect{The mbTLS Protocol}
The mbTLS protocol~\cite{mbtls} is a proposed middlebox-enabled TLS
scheme. It differs from standard TLS in the following ways:

\begin{itemize}
    \item During the mbTLS handshake between
    a client and a server, middleboxes
    report their presence to the endpoints
    with an additional handshake.
    \item At the end of the handshake phase,
    each middlebox is delegated two session keys
    for its associated sessions
    (i.e. one where it acts as a client, and one as a
    server)
    from the endpoint which has deployed them.
    \item During the record phase, messages between the
    endpoints are passed down the chain of middleboxes,
    which decrypt then re-encrypt each message 
    (whilst performing any of their functions on the message
    body)
\end{itemize}

Importantly, mbTLS only proposes significant alterations to the handshake
phase of the TLS protocol.
Once the session has been established, each segment in the path performs the TLS record phase protocol as usual. This is in contrast to e.g. mcTLS~\cite{mctls}, in which agents who perform modifications re-encrypt the payload with a different key, or maTLS~\cite{matls}, where a MAC is appended to the message at each segment. An overview of the mbTLS scheme is given in
Figure~\ref{fig:mbtls-skipping-attack}. 

The mbTLS scheme assumes that middleboxes are running on hardware
enclaves, such as the Intel SGX framework~\cite{sgx}. As a result,
the authors assert that middleboxes can be seen as trusted agents for
the purpose of security analysis. However, we argue that this is
not a realistic security model. Software enclaves are designed
to ensure that software is loaded without modifications
after distribution, and is run in a secure environment.
This does not provide any guarantees about security
of the software itself.

As such, maliciously designed or configured
middlebox software is not protected against.
This means that
an attacker could write malware which poses as a service-providing
middlebox (such as an ad-blocker), but instead fulfills some other
purpose (such as leaking data). In addition, we note the
existence of several attacks on trusted execution 
environments~\cite{foreshadow, cache, stealthy, resolution},
suggesting that even well-intending participants may accidentally
leak secret data.

\mysubsect{Middlebox Skipping Attack on mbTLS}
Our analysis shows that the mbTLS protocol admits a skipping attack. 

\begin{figure}[h]
    \centering
    \captionsetup{justification=centering}
    \includegraphics[width=0.45\textwidth]{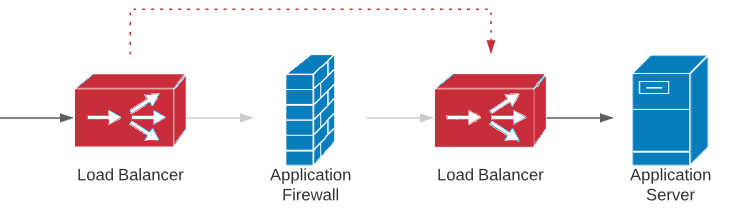}
    \caption{Firewall is bypassed by two collaborating agents}
    \label{fig:mbtls-attack-explained}
\end{figure}

Looking back to Figure~\ref{fig:mbtls-network-structure}, suppose
a malicious administrator has access to the two layers of load
balancers. During the mbTLS handshake
phase, the Application Firewall is registered
as being an active participant in the session.
However, once packets are sent during the record phase,
the two load balancers communicate directly through a
side channel, bypassing the application
firewall. This idea is shown in Figure~\ref{fig:mbtls-attack-explained}.

The application server is unable to
differentiate between the firewall choosing not to modify a
message, and the firewall never having received the message at all.
In this way, malicious inputs could be targeted against the AS
without the protection of the firewall.

Intuitively, the Application Server
should have some guarantee that because it saw the firewall
registered during the handshake phase, the firewall continues to
be an activate participant in all record phase messages.
This leads to the following requirement on message-forwarding
protocols:
\begin{defn}{Path Integrity (Notion)}
\label{def:path-integrity-idea}
    Once a path has been established, if a message
    is received by one agent on a path, then all
    previous agents on the path should have also received it.
\end{defn}
In order to prevent an attack such as the one here,
a simple approach is to add some form of read-receipt to messages,
in the form of a MAC or signature from each middlebox in the path.
Upon receipt of a message, endpoints can
confirm that the path was followed faithfully by checking that the
set of signatures has been correctly constructed.
Indeed, the mcTLS~\cite{mctls} scheme makes use of
``write keys'' that privileged middleboxes use to update a write MAC, but possesses a similar vulnerability to mbTLS in that if no
changes are made to the payload, the associated message
also remains unchanged outside re-encryption.


%

%

\section{Multiset Rewriting Theory}
\label{sec:msr}

In this section, we introduce the framework that will be used to
describe protocol execution and analysis for the rest of the paper. The
language we use can be considered as a subset of that supported by the
Tamarin prover tool~\cite{tamarin,tamarinmanual}.


We employ a multiset rewriting system -- a special form of
term rewriting system. The state of the communication network is
modelled by a collection of \textit{facts}, with rewrite rules
which add or remove from this collection, forming a labelled transition
system.


\subsection{Fundamentals}

Our term rewriting system is built using terms from an
order-sorted algebra (e.g. Goguen~\cite{orderSorted}).
We define two top-level sorts $\type{msg}$ and $\type{Fact}$, and subsorts
$\type{pub}$, $\type{fresh}$, such that $\type{pub} < \type{msg}$
and $\type{fresh} < \type{msg}$.

Intuitively, $\type{msg}$ terms represent any value that might
be used on the communication network, while $\type{pub}$ and
$\type{fresh}$ terms represent public and freshly generated values,
respectively.
We write $x\colon\type{y}$ to indicate that
the term $x$ is of type $y$.
We define a public term 
$\text{`'}\colon\type{pub}$, representing the empty string.
Public terms will also be used to indicate agents' identities.
Notationally, we write the following symbols
to indicate the type of a term:
$A \colon\type{pub}$,
    $x \colon\type{fresh}$,
    $m \colon\type{msg}$.



We allow for collections of function symbols $\Sigma_{\type{msg}^\ast, \type{msg}}$ 
and $\Sigma_{\type{msg}^\ast, \type{Fact}}$ , which map a sequence of type $\type{msg}$
(or its subsorts) to either another $\type{msg}$ or a $\type{Fact}$. For simplicity
we denote $\Sigma_{\type{msg}^\ast, \type{msg}}$ by $\functionSet$ and 
$\Sigma_{\type{msg}^\ast, \type{Fact}}$ by $\factSet$.

Atoms (i.e. undecomposable terms) can represent names (i.e. unassigned expressions),
or variables (i.e. assigned values).
A term is said to be \textit{ground} if it contains no variables.
A \textit{substitution}, $\sigma$, is a (partial) function from variables to terms of the same sort
(or a subsort). We say a substitution is \textit{ground} if it
maps to a ground term. 
A substitution is applied to a term by applying it to each subterm. Given a term $t$ and
a substitution $\sigma$, we say that the substitution $\sigma$ \textit{grounds} $t$ if the resultant term
$t\sigma$ is a ground term. 
We allow for an equational theory $E$ over terms of type $\type{msg}$, or its subtypes.
$E$ is a collection of equations $\mi{lhs} = \mi{rhs}$.
Two terms $s$ and $t$
are said to be equivalent modulo $E$ if a series of equations can be applied
to $s$ or $t$ (or both) such that the resulting terms $s^\prime$ and $t^\prime$
are equal.
We will make use of a standard set of function
symbols $\functionSet$,
as well as an associated
equational theory $E$, which is presented in
Table~\ref{table:equational-theory}.

\begin{table}[h]
    \normalsize
    \begin{tcolorbox}
    \centering
    \begin{tabular}{p{1.4cm}  p{5.5cm}}
        \multicolumn{1}{c}{\bf Symbols} & \multicolumn{1}{l}{\bf Equations} \\
        \hline
        & Hashing \\
        \multicolumn{1}{c}{$\h/1$} & \\
        \hline
        & Pairing \\
        \multicolumn{1}{c}{$\langle \rangle/2$} & \\
        \multicolumn{1}{c}{$\fst/1$} & $\fst(\langle x, y \rangle) = x$ \\
        \multicolumn{1}{c}{$\snd/1$} & $\snd(\langle x, y \rangle) = y$ \\
        \hline
        & Symmetric encryption \\
        \multicolumn{1}{c}{$\senc/2$} & \\
        \multicolumn{1}{c}{$\sdec/2$} & $\sdec(\senc(m, k), k) = m$ \\
        \hline
        & Asymmetric encryption \\
        \multicolumn{1}{c}{$\pk/1$} & \\
        \multicolumn{1}{c}{$\aenc/2$} & \\
        \multicolumn{1}{c}{$\adec/2$} & $\adec(\aenc(m, \pk(k)), k) = m$ \\
        \hline
        & Signatures \\
        \multicolumn{1}{c}{$\sign/2$} & \\
        \multicolumn{1}{c}{$\true/0$} & \\
        \multicolumn{1}{c}{$\verify/3$} & $\verify(\sign(m, k), m, \pk(k)) = \true$
    \end{tabular}
    \end{tcolorbox}
    \caption{Collection of functions $\Sigma$ and
    equational theory $E$}
    \label{table:equational-theory}
\end{table}

We write $\{ x \}_k$ to denote encryption when the
kind is clear from the context. 
Similarly, we will sometimes omit
the pair operators $\langle$ and $\rangle$ for readability.


\subsection{Protocol Specification}

\DONE{Describe the basics of a multiset rewriting system}

Thusfar, our discussion has been restricted to
terms of type $\type{msg}$. We now divert our
attention to \textit{Facts}. Intuitively, while
$\type{msg}$ terms model the value of certain things
(messages, agent names, encryption keys), terms of
type $\type{Fact}$ describe the state of the protocol
execution itself. As such, facts are often parameterised
by one or more message terms.
We reserve the following fact symbols with
corresponding intuition:

\begin{itemize}
    \item $\factName{Net}/1$: A message on the communication network
    \item $\factName{K}/1$: Adversary knowledge of a term
    \item $\factName{Pk}/2, \factName{Ltk}/2$: Public and long-term keys
    \item $\factName{ShKey}/3$: A shared encryption key
\end{itemize}

From now on we assume that
we are working over multisets where all terms are of type $\type{Fact}$. A
\textit{State}, $S$, is a multiset where all of the terms are ground, and
models the current execution state of a protocol. Each run of a protocol will begin with an empty state, which
then transitions into future states through a series of \textit{rules}.

\DONE{Define a state as a ground multiset of facts}

A \textit{rule} $r$ is defined by a triplet of multisets
$r\colon L \xrightarrow{E} R$. Given a
state $S$, and a substitution $\sigma$, we can apply rule $r$ if:
\begin{itemize}
    \item $\sigma$ is a grounding substitution for $L$ and $R$, and
    \item $L\sigma \subset S$.
\end{itemize}

In this case, the state $S^\prime$ is produced by removing the submultiset of
$S$ equal to $L\sigma$, and replacing it with $R\sigma$.
The elements of $E\sigma$
are known as the \textit{event facts} of the rule.

\DONE{Define persistent fact symbol notation for convenience}

For convenience, we will sometimes use the prefix $!$ in fact symbols
to indicate that the fact is \textit{persistent}. Persistent facts are
never removed from a state as a result of rule execution
. They represent reusable assets, such as encryption keys or
adversary knowledge.

A simple example of a rule is given in
$\ruleLabelFont{Dec\_Fwd}$ below, in which an agent
(whose name will instantiate the variable $A$),
detects a message
on the network encrypted with their public key, and decrypts it
before forwarding on the result.

\begin{align*}
\makeNamedRule{
    Dec\_Fwd
}{
    \ruleFact{Net}{\{m \}_{pk(k)}} \\
    \ruleFact{!Ltk}{A, k}
}{
    \ruleFact{Fwd}{A, m}
}{
    \ruleFact{Net}{m}
}
\end{align*}

When a rule is applied, the terms $E\sigma$ are appended to the \textit{trace}, $\tau$, an indelible ordered history of event markers.
At the start of any execution, $\tau = \phi$, the empty trace. After the
execution of a rule $r$, the resulting event facts are added along with
a discrete time marker $\hash t_i$. For example, an application of
the rule above might append the fact 
$\ruleFactAtTime{Fwd}{A, m}{t_1}$ to the trace.
Time markers are assumed to be ordered, unique, and increasing. However, they hold no values (they do not
represent actual timestamps, only the order of events in an 
execution).

We will freely quantify over time markers e.g. $\forall \hash t_i$ when there
is no ambiguity, and will make use of ordering of time markers
(e.g. $\hash t_i < \hash t_j$). 

We reserve the following special rule $\ruleLabelFont{Fresh}$:
\begin{align*}
\makeNamedRule{
    Fresh
}{
\hspace{2ex}-\hspace{2ex}
\ }{
\hspace{4ex}
}{
    \ruleFact{Fr}{x:\type{fresh}}%
}
\end{align*}
This rule allows for the creation of freshly generated random
variables, for example for use in creating encryption keys. We specially require
that each execution of the
$\ruleLabelFont{Fresh}$ rule is instantiated
by a distinct, previously unused value for $x$, and that it is the only rule which
can create $\factName{Fr}$ facts.
Since the created $\factName{Fr}$ fact
is linear, it is consumed if used by
a later rule -- this ensures that the same
random value can never be generated twice.

\DONE{Describe how the Fresh rule impacts execution, and
the deducibility of fresh values}

\DONE{Define a ``Protocol'' as a collection of rules with
some additional baggage (function symbols etc.)}

A protocol, $\protocol$, is given by $\protocol = (\ruleSet, \factSet, \functionSet, E)$,
a tuple of rules, facts, functions and an equational theory. We will
assume that $\ruleSet$, $\factSet$, $\functionSet$ and $E$ contain all
the reserved elements indicated in this section (including the adversary
rules below). 
We define $\allTraces{\protocol}$ as the set of all
(valid) traces that can be constructed as a result of
executing the rules from $\ruleSet$ along with the associated
material from the other protocol components.

\subsection{Adversary Model}
\label{subsec:adversary}

Our model uses the Dolev-Yao~\cite{dolevyao} adversary.
It is also defined in terms of multiset
rewriting rules, given in Figure~\ref{table:adversary-rules}.
The Dolev-Yao adversary is capable of eavesdropping, modifying,
and retransmitting messages, modelled
by the $\ruleLabelFont{Block}$ and $\ruleLabelFont{Inject}$
rules. The 
$\ruleLabelFont{Adv\_Pub}$ and $\ruleLabelFont{Adv\_Fr}$ rules
allow the adversary to deduce public and
(previously unused) fresh terms, while 
$\ruleLabelFont{Fun_f}$ allows the adversary
to derive new terms by applying function symbols to known terms.
Finally, the various $\ruleLabelFont{Corrupt}$
rules model
agents who are fully under the control of the adversary.


\begin{figure}[ht]
\begin{align*}&
\makeNamedRule{
Inject
}{
    \ruleFact{!K}{x}
}{
}{
    \ruleFact{Net}{x}%
}
\\&
\makeNamedRule{
Block
}{
    \ruleFact{Net}{x}
}{
    \ruleFact{K}{x}
}{
    \ruleFact{!K}{x}%
}
\\
\\&
\makeNamedRule{
    Adv\_Pub
}{
\hspace{2ex}-\hspace{2ex}\ 
}{
    \ruleFact{K}{A}
}{
    \ruleFact{!K}{A : \type{pub}}%
}
\\&
\makeNamedRule{
Adv\_Fr
}{
    \ruleFact{Fr}{x}
}{
    \ruleFact{K}{x}
}{
    \ruleFact{!K}{x}%
}
\\&
\makeNamedRule{
    Fun_f
}{
    \ruleFact{!K}{x_1}\\
    \;\;\, \cdots\\
    \ruleFact{!K}{x_2}
}{
    \ruleFact{K}{f(x_1, \ldots, x_n)}
}{
    \ruleFact{!K}{f(x_1, \ldots, x_n)}%
}
\\ 
\\&
\makeNamedRule{
    Corrupt\_Ltk
}{
    \ruleFact{!Ltk}{A, k}
}{
    \ruleFact{Corrupt}{A}
}{
    \ruleFact{!K}{k}\;%
}%
\\&
\makeNamedRule{
    Corrupt\_L
}{
    \ruleFact{!ShKey}{A, B, k}
}{
    \ruleFact{Corrupt}{A}
}{
    \ruleFact{!K}{k}%
}%
\\&
\makeNamedRule{
    Corrupt\_R
}{
    \ruleFact{!ShKey}{A, B, k}
}{
    \ruleFact{Corrupt}{B}
}{
    \ruleFact{!K}{k}%
}%
\end{align*}
\caption{Rules which define the Dolev-Yao adversary.
\label{table:adversary-rules}%
    }
\end{figure}


\subsection{Security Properties}

\DONE{First order logic claims over our execution model}

Security goals of a protocol are given in terms of first-order logic
formulae on the set of traces of the protocol. Intuitively, they
indicate that certain events can or cannot happen, or that they
must occur in a certain order.

A security goal holds for a given protocol if all
traces of the protocol satisfy it. A trace which
violates the goal can be reconstructed into an
attack on the protocol.
As a simple example, consider the following protocol $P$ in
Alice-Bob notation:
\begin{align*}
     A \xrightarrow{} B: \qquad & \{ x, y \}_{k_{AB}}  \\ 
     B \xrightarrow{} A: \qquad & \{ y, x \}_{k_{AB}}
\end{align*}
In this protocol, $A$ sends a pair of encrypted terms to $B$, who
reverses their order and sends them back. A reasonable security
claim for this protocol might be ``The value $x$ is either unknown
to the adversary, or one of $A$ or $B$ is corrupt".
This could be expressed as:
\begin{align*} &
    \forall A, B, x, t_i\colon
    \ruleFactAtTime{Secret}{A, B, x}{t_i} \implies \\ &
\hspace{2ex}
    \not\exists \hash t_a\colon  \ruleFactAtTime{K}{x}{t_a} \logicOr 
    \exists \hash t_b\colon  \ruleFactAtTime{Corrupt}{A}{t_b} \logicOr 
    \exists \hash t_c\colon  \ruleFactAtTime{Corrupt}{B}{t_c}
\end{align*}

Note that this definition implies that the adversary can
\textit{never}
learn the value $x$. We can add event orderings (such as
$\hash t_a < \hash t_i$) to refine these claims. Our
claims are generally quantified over all traces
$\tau \in \allTraces{P}$,


We also use the same syntax to restrict the set of traces of
a protocol to be investigated. As an example, we introduce the
\factName{Equal} fact, denoting a test for equality, 
and consider only traces where this test is successful:
%
$
    \forall x, y, \hash t_i \colon
    \ruleFactAtTime{Equal}{x, y}{t_i} \implies x = y 
$.

Such a restriction ensures we only consider
traces in which
signatures are correctly verified -- for example,
a rule containing 
$\ruleFact{Equal}{\verify(sig, msg, pkA), \true}$
can only be executed in the case that the value $sig$
is indeed a signature for $msg$.

\section{Modelling Path Protocols}
\label{sec:modelling-path-protocols}

In this section we introduce the notion of a \textit{path protocol}.
We break down the structure of a path protocol into
a set of phases, and describe each of these phases
as a collection of generic rules.

\subsection{Running Example}

\begin{figure}[t]
    \centering
    \includegraphics[width=0.45\textwidth]{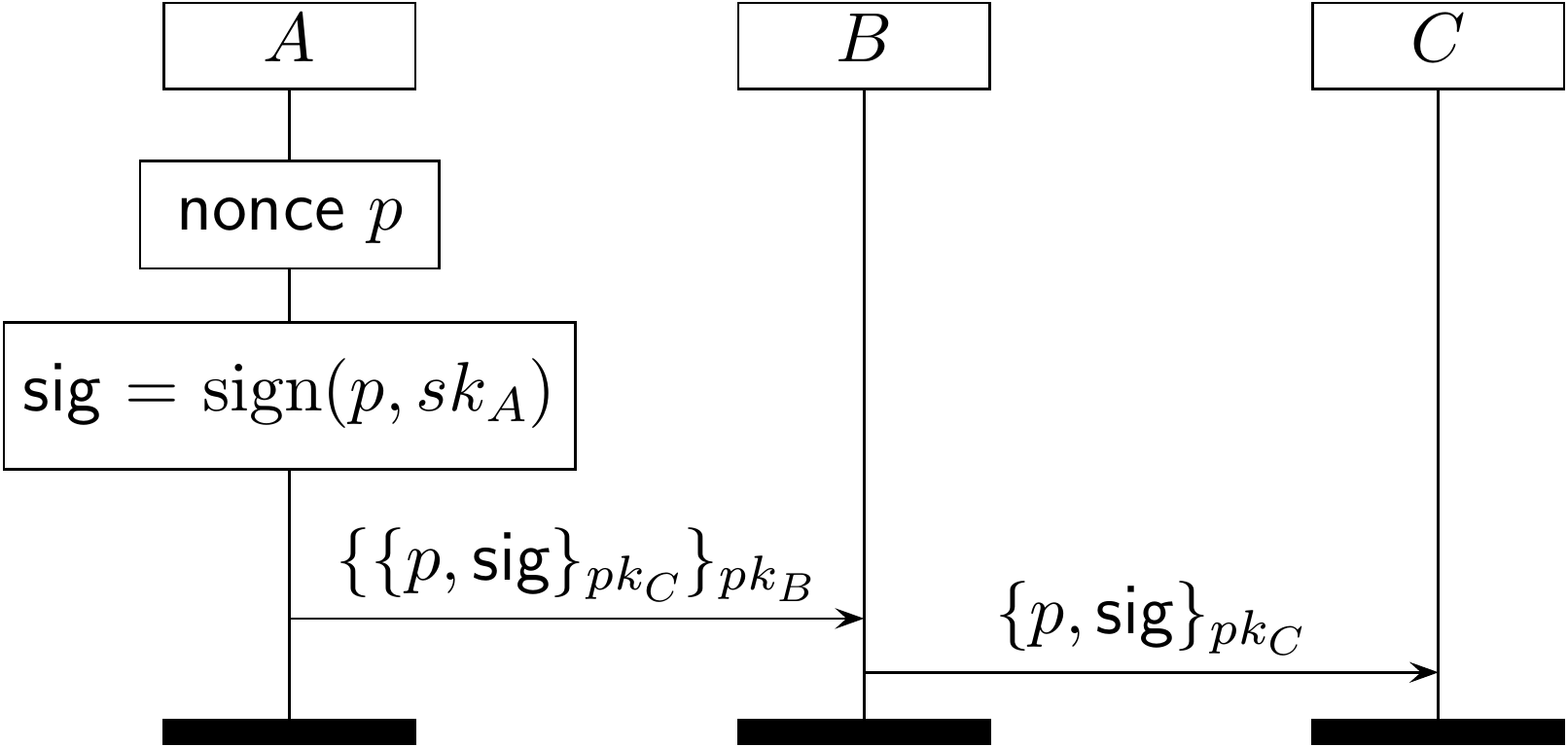}
    \caption{A simple message forwarding protocol.%
    \label{fig:public-key-oneway}}
\end{figure}

Consider the multi-party message forwarding protocol shown in
Figure~\ref{fig:public-key-oneway}, which uses public key encryption.

Intuitively, the agent $A$ wants an intermediate agent $B$
to forward a message $p$ to $C$.
This is achieved by using nested encryption.
The protocol is depicted with one forwarding agent $B$, but
indeed it could be trivially extended for any number
of forwarding agents.

It would be relatively straightforward to 
model this protocol for a fixed number of
agents, by specifying the value of the message at each
step of execution.
However, this design quickly becomes cumbersome as the
number of agents grow. Moreover, it requires a separate model for
each number of agents. Instead, we construct
a single model which accounts for any number of
agents.
Figure~\ref{fig:message-rolling-rules} shows the
set of rewriting rules which model this protocol.

\subsection{Modelling Multi-Step Messages}
Our model in Section~\ref{sec:msr} considers messages to be sent between
pairs of agents. We now build a set of rewriting rules which allow us to
specify protocols that use these multi-step messages. Intuitively, such
messages are constructed by ``wrapping'' layers of encryption on top of
each other. This approach will allow us to define \textit{path protocols},
which we break down into a series of phases.

\begin{defn}[Path Protocols (Notion)]
\label{def:PathProtocols-mk1}
A \textit{Path Protocol} is a protocol in which rules
(other than those modelling adversary
capabilities) can be categorised as belonging to one of
the following phases:
\begin{itemize}
    \item \textbf{Setup Phase}: A preliminary phase in which
    agents' encryption keys are established
    \item \textbf{Construction Phase}: An initial
    agent creates a message from a payload, by configuring
    it to pass through a series of intermediate agents
    \item \textbf{Forwarding Phase}: Each intermediate
    agent receives, repackages and forwards the message
    \item \textbf{Receive Phase}: The intended recipient
    receives the final message and retrieves the payload.
\end{itemize}
\end{defn}
\DONE{Define construction phase, payload, message}
Throughout the course of our discussion, we will generally
use the name $p$ to refer to the \textit{payload} -- the
intended value for the final recipient. The
name $m$ will be used to refer to \textit{messages} sent
between agents -- including things like encryption (which
we do model) and header or miscellaneous data (which we
do not).
We make the assumption that for each
individual session, the payload will always
contain some session data or randomness that
makes it unique, and thus suitable as a
session identifier.

Over the course of this section, we 
build a framework 
of generic rules which is sufficient to cover each of the individual phases
in Definition~\ref{def:PathProtocols-mk1}. This resulting
set of rules can be used to describe a large majority of path
protocols.


\DONE{Informal definition of path order}

\begin{figure}[t]
\begin{flushleft}
\textbf{Setup Phase}
\end{flushleft}
\begin{align*}
&\makeNamedRule{ 
Gen\_Ltk
}{
    \ruleFact{Fr}{\ltk}
}{
    \hspace{4ex}
}{
    \ruleFact{!Ltk}{A, \ltk} \\%
    \ruleFact{!Pk}{A, \pk(\ltk)}\\
    \ruleFact{Net}{\pk(\ltk)}
}
\hspace{-3em}\xdash[1em]
\hspace*{12cm} 
\end{align*}
\begin{flushleft}
\textbf{Construction Phase}
\end{flushleft}
\begin{align*}
&\makeNamedRule{
Create
}{%
    \ruleFact{Fr}{p}\\
    \ruleFact{!Pk}{E, pkE}\\
    \ruleFact{!Ltk}{A, ltkA}
}{
    \ruleFact{Add}{p, E, \{ p, m , '\ '}\\
    \ruleFact{StartBuild}{A, p}
}{
    \ruleFact{Build}{p, E, m}%
}%
\hspace*{12cm} 
\\[0.8em]& \qquad \qquad m = \{ p, \sign(p, ltkA) \}_{pkE} \hspace*{12cm}
\\[0.8em]%
&\makeNamedRule{
Wrap
}{
    \ruleFact{Build}{ p, M_i, m}\\
    \ruleFact{!Pk}{ M_{j}, \mi{pk} }
}{
    \ruleFact{Add}{p, E, m, \{ m \}_{\mi{pk}} }
}{
    \ruleFact{Build}{p, M_{j}, \{ m \}_{\mi{pk}}}%
} %
\hspace*{12cm} 
\\[0.8em]
&\makeNamedRule{
Send
}{
    \ruleFact{Build}{p, M_1, m}
}{
    \hspace{4ex}
}{
    \ruleFact{Net}{m}%
}%
\hspace*{12cm} 
\end{align*}
\begin{flushleft}
\textbf{Forwarding Phase}
\end{flushleft}
\begin{align*}
& \makeNamedRule{
Unwrap
}{
    \ruleFact{Net}{ \{ m \}_{\pk(\ltk)} }\\
    \ruleFact{!Ltk}{M_i, \ltk}
}{
    \ruleFact{Forward}{M_i, \{ m \}_{\pk(\ltk)}, m }
}{
    \ruleFact{Net}{m}%
}%
\hspace*{12cm} 
\end{align*}
\begin{flushleft}
\textbf{Receive Phase}
\end{flushleft}
\begin{align*}
&\makeNamedRule{
Receive
}{
    \ruleFact{Net}{ \{ p, sig \}_{\pk(ltkE)} }\\
    \ruleFact{!Ltk}{E, ltkE)}\\
    \ruleFact{!Pk}{A, pkA}
}{
    \ruleFact{Forward}{E, \{ p, sig \}_{\pk(ltkE)}, '\ ' }\\
    \ruleFact{Equal}{\verify(sig, p, pkA), \true}
}{
-
}%
\hspace*{12cm} 
\end{align*}
    \caption{Full set of rewriting rules for the example protocol given in
    Figure~\ref{fig:public-key-oneway}.
        \label{fig:message-rolling-rules}}
\end{figure}

\mysubsect{Setup Phase}
Our execution model begins with the empty multiset.
In order for the protocol to begin, agents must be
instantiated and assigned asymmetric and shared encryption
keys.
Our models make use of
the $\ruleLabelFont{Gen\_ShKey}$ and
$\ruleLabelFont{Gen\_Ltk}$ rules for generating
encryption keys, specified as follows:
\begin{align*}
&\makeNamedRule{ 
Gen\_ShKey
}{
    \ruleFact{Fr}{k}
}{
    \hspace{4ex}
}{
    \ruleFact{!ShKey}{A, B, k}%
}%
\hspace*{12cm} 
\\ %
&\makeNamedRule{ 
Gen\_Ltk
}{
    \ruleFact{Fr}{\ltk}
}{
    \hspace{4ex}
}{
    \ruleFact{!Ltk}{A, \ltk} \\%
    \ruleFact{!Pk}{A, \pk(\ltk)}\\
    \ruleFact{Net}{\pk(\ltk)}
}
\end{align*}

We also allow for the corruption of agents created
during the setup phase.
We will
assume that the initiating agent in each session is honest,
but allow all other agents to be under full adversarial control. 

\mysubsect{Construction Phase}
The Construction Phase represents the beginning of a session
of the protocol. We break this down into three rules: a
$\ruleLabelFont{Create}$ rule which determines the payload,
a $\ruleLabelFont{Wrap}$ rule which modifies the message to
pass through an intermediate agent, and a 
$\ruleLabelFont{Send}$ rule, in which the message is released
onto the network.

Implicit in the application of these rules is the notion of
a \textit{path order}. Intuitively, this is a relation that
models the (intended) order in which the message will pass
between protocol participants.
The construction phase that takes place in
each run of the protocol
will define the path order for that
execution. 


\begin{defn}[Path Order]
    \label{defn:path-order}
    A Path Order is a total order $<_\pi$ on a
    (finite) set of public terms.
    We call the minimal element $A$ of $<_\pi$ the
    \textit{initial} agent, the maximal element $E$ the
    \textit{final} agent, and all other elements $M_i$
    \textit{intermediate} agents.
    \DONE{A Path Order is...}
\end{defn}

We allow for a different 
path to be chosen for each session.
We assume that our protocols are
designed such that paths are non-repeating
(i.e. the path
order is always well-defined).
A path order need not include all agent identifiers,
and so the path can be of any length.
The idea of a path order allows us to
present an informal notion for 
the Path Integrity security goal that we will define
in the next section.


\begin{defn}[Path Integrity (Revisited)]
\label{def:path-integrity-informal}
    A protocol satisfies path integrity if for every session,
    for all $M_i$ and $M_j$ such that $M_i <_\pi M_j$,
    if $M_j$ has forwarded the message, then
    $M_i$ has also forwarded the message.
\end{defn}


In order to formalise this notion, we will make use of
\textit{expected messages}. We define the
linear fact \ruleFact{Build}{p, M, msg},
which represents the
message as it is being constructed: the payload $p$
is used as a unique path identifier, while the second and
third terms indicate the current agent being considered
as well as the current value of the message (for example,
as
successive layers of cryptography are applied).
For more complex protocols, additional parameters may be
added to the \factName{Build} fact to track state.
The event fact \factName{Add} represents that
an agent has been added
to the path. It is parameterised by the
path identifier, the agent who has been
added to the path, and how the initiating
party anticipates the message will be
altered as it passes through them (for example, through
de- or re-encryption).
The \factName{StartBuild} event fact is used to mark the
beginning of the protocol execution.

These facts are used by three rules during this phase. In
the first rule, the initiating agent determines the payload
to be sent to the other endpoint.
The second rule is repeatedly applied to add new intermediate
agents to the path in order, each time replacing the current
\factName{Build}
fact with a new version containing any required changes to the message. Finally, the last rule sends the 
message on to the network:
\begin{align*}
&\makeRule{
    \ruleFact{Fr}{p}\\
    \ruleFact{!Pk}{E, pkE}
}{
    \ruleFact{Add}{p, E, f(p), \text{`'}}\\
    \ruleFact{StartBuild}{A, p}
}{
    \ruleFact{Build}{p, E, f(p)}%
}\\
&\makeRule{
    \ruleFact{Build}{p, M_i, m}\\
    \ruleFact{!Pk}{ M_{j}, pkJ }
}{
    \ruleFact{Add}{p, M_i, m, g(m)}
}{
    \ruleFact{Build}{p, M_{j}, g(m)}%
}\\
&\makeRule{
    \ruleFact{Build}{p, M_1, m}
}{
    \hspace{4ex}
}{
    \ruleFact{Net}{m}%
}%
\end{align*}
We use anonymous functions $f$ and $g$ to depict how
the message is changed -- these are instantiated for each
specification based on the protocol in question.
For our example protocol, $f$ is pairing
with a
signature, and $g$ is asymmetric encryption using a
public key.

The primary requirement of our ``wrapping" function $g$ (and later on
the associated ``unwrapping" function) is that the general structure
of the message is preserved as it is transmitted between agents.
In our example, each agent expects to receive (and send) a message
that consists of a single term encrypted by a public key, $\{ m \}_{pk}$.
In a more complex situation, this message may contain components such as
a public term (containing the identity of the next agent in the path), or
a list of signatures. We make this requirement on the protocol specification
level -- an agent may not be able to verify themselves that part of the encrypted
body of their message has this structure. We require this structural symmetry
between the four following message types sent during protocol execution:

\begin{itemize}
    \item The outwards packet sent by $A$
    \item The final packet received by $E$
    \item Each packet going into a forwarding agent $M_i$
    \item Each packet going out of a forwarding agent $M_i$
\end{itemize}

The
ordering of \factName{Add} facts in any
given trace establishes a path
order $<_\pi$: If the \factName{Add} fact
for $M_i$ is added to the trace before that of
$M_j$, we have that
$M_j <_\pi M_i$.

\mysubsect{Forwarding Phase}
The forwarding phase of the protocol occurs as intermediate
agents transceive the message
We model this with the use of the
\ruleLabelFont{Unwrap} rule, in which each
intermediate agent forwards the message.
The exact nature of this forwarding is dependent on the protocol -- it may involve de- or re-encryption, or reading
information about how to route the forwarded message.
%
\begin{align*}
\makeRule{
    \ruleFact{Net}{ m }\\
    \ruleFact{!Ltk}{M_i, \ltk}
}{
    \ruleFact{Forward}{M_i, m, f(m) }
}{
    \ruleFact{Net}{f(m)}%
}%
\end{align*}

We introduce the \factName{Forward} fact to denote that the
agent has forwarded the message, including the values it has
changed from and to. In a faithful execution
of the protocol,
the parameters of these facts should agree with
those in the \factName{Add}
facts created in the Construction phase.

\mysubsect{Receive Phase}
The last step of a protocol is upon the successful receipt
of the payload by the endpoint.
\begin{align*}
\makeRule{
    \ruleFact{Net}{ f(p) }\\
    \ruleFact{!Ltk}{E, \mi{ltkE})}\\
}{
    \ruleFact{Forward}{E, f(p), \text{`'} }
}{
\hspace{2ex}-\hspace{2ex}\ 
}%
\end{align*}
 
The \ruleLabelFont{Receive} rule may include
validation of the final
payload. For example, in the example protocol,
the final agent must ensure that the attached
signature matches the payload. More advanced
validation is discussed in the following
section.
%

\begin{figure}[b]
\begin{tcolorbox}
\vspace{-2ex}
\[{\setstretch{1.2} \begin{array}{ r p{0.5cm} l p{0.5cm} p{12cm}}
 && \forall A, M_i, M_j, p_{ID}, f_i, t_i, f_j, t_j, &%
\\ &&  \hspace{4ex} \hash ta_i, \hash ta_j, \hash tk_i, \hash ts \st
&&
\\ 
1. && (\not\exists \hash ta_c \st \ruleFactAtTime{Corrupt}{A}{ta_c})
\\ 
2. && \hspace{0ex} \wedge \ruleFactAtTime{StartBuild}{A, p_{ID}}{ts}
\\ 
3. & &\hspace{0ex} \wedge \ruleFactAtTime{Add}{p_{ID}, M_i, f_i, t_i}{ta_i}
\\
4. && \hspace{0ex} \wedge \ruleFactAtTime{Add}{p_{ID}, M_j, f_j, t_j}{ta_j}
\\
5. && \hspace{0ex} \wedge (\hash ts < \hash ta_i < \hash ta_j)
\\
6. && \hspace{0ex} \wedge \big(
    \ruleFactAtTime{Forward}{M_i, f_i, t_i}{tk_i} \big) 
\\
7. && \implies
\\ 
8. && \hspace{0ex} \exists \hash tk_j \st \, (\hash tk_j < \hash tk_i)
\\ 
9. && \hspace{0ex} \wedge \big(
\\ 
10.  && \hspace{4ex} \big( \ruleFactAtTime{Forward}{M_j, f_j, t_j}{tk_j} \big)
\\ 
11. && \hspace{2ex} \vee \, \big( \, \exists\  \hash tc_j \st \ruleFactAtTime{Corrupt}{M_j}{tc_j} \land
\\ 
12. && \hspace{6ex} \ruleFactAtTime{K}{\langle f_j, t_j \rangle}{tk_j}  \big)
\\ 
&& \hspace{2.15ex} \, \big)%
\end{array}
}\]
\end{tcolorbox}
\caption{Statement of the Path Integrity security goal}
\label{fig:path-integrity-equation}
\end{figure}
\section{Security Goals for Path Protocols}
\label{sec:path-integrity}

We introduce two security goals to cover the range of
protocols built in the framework from the previous section.
The first,
\textit{Path Integrity} (Sec.~\ref{subsec:path-integrity}),
covers the simplest case of message forwarding protocols. 
The second, \emph{Verification-dependent Path Integrity}
(Sec.~\ref{subsec:verification-dependent-path-integrity})
makes Path Integrity conditional on the receiving
party validating the received message.
Note that these security goals are perpendicular to many existing security
goals, e.g. those relating to secrecy or synchronisation between agents. For example,
if an attacker is able to confuse an agent into believing that they are
performing a different role in the protocol (e.g. that they are an endpoint when
they should be forwarding the message), other attacks may arise. The
security goals we present here form an extension of the classical security
goals for this protocol domain, not a replacement.


\mysubsect{Intuition}
\label{subsec:intuition}
The intuition behind the structure of our security
goals is as follows. Given a specific protocol session,
we assume that a path order $<_\pi$ has been defined by a
sequence of events. We then examine the case that some
agent $M_i$ has successfully forwarded the message.
Our goal is satisfied if there is (fundamentally)
only one way to fill the gap between these
events: that each intermediate
agent $M_j$ such that $M_j <_\pi M_i$ has also
forwarded the message. These claims are typically verified
by performing backwards reasoning -- checking all ways of
reconstructing each partial trace.


\subsection{Path Integrity}
\label{subsec:path-integrity}

We begin by formalising
Definition~\ref{def:path-integrity-informal}, the
idea of \textit{Path Integrity}. This goal represents the
initiating agent's belief that a sent message will indeed
travel through the list of intermediate agents in the
intended path in the correct order.
Intuitively, this requires a correspondence between the
order in which agents were named in applications of the
\ruleLabelFont{Wrap} and \ruleLabelFont{Unwrap} rules.

\begin{defn}{Path Integrity}
\label{def:one-way-path-integrity}

\DONE{We say that a protocol satisfies one-way path integrity if...}

We say that a protocol $P$ satisfies Path Integrity if and only if all traces $\tau \in \allTraces{P}$ satisfy
the property displayed in
Figure~\ref{fig:path-integrity-equation}.

\end{defn}

The intuition of the definition in Figure~\ref{fig:path-integrity-equation} 
is as follows: (1) Suppose $A$ is an honest agent, who (2) starts a session ID $p_{ID}$,
(3) adding agents $M_I$ and (4) $M_j$ to the path, such that (5) $M_j <_\pi M_i$,
and (6) $M_i$ has successfully forwarded the message, (7) then (8) at some earlier
time, (9) either: (10) $M_j$ forwarded the message, (11) or $M_j$ is corrupt and
(12) the adversary had the necessary knowledge to forward the message.

Note that in case the agent is corrupt, the adversary may forward the message.
In this a situation, the path order is still
preserved, as the message was still correctly forwarded.

\subsection{Verification-Dependent Path Integrity}
\label{subsec:verification-dependent-path-integrity}

The security property defined in the previous section
can be seen as \textit{on-the-fly} security:
it ensures that path integrity holds even while a
message is in-flight. This is common for
onion-style protocols, which use layered
encryption such that the message can
only be decrypted in a specific order.
However, for many protocols,
this requirement might be too strict -- we may
want to loosen it to only consider
\textit{completed} sessions.
This approach is typical in middlebox-enabled TLS protocols~\cite{mctls, matls, metls}, where
an additional \textit{verification phase} exists to ensure that a
session has executed completely. We extend our framework 
to model such protocols and their security.



\mysubsect{Verification Phase}
We assume that the final message received by the endpoint can be
broken into two main parts: one containing the session payload
(or some function thereof), and another which includes validation
data, such as appended MACs.

We modify the \ruleLabelFont{Receive} rule to
include a
$\ruleFact{Check}{E, f(p), m}$ fact, separating
these two components. Successive applications of
a \ruleLabelFont{Verify} rule then check the
validation data added by each
intermediate agent in turn. A final rule runs
after all verification steps to confirm
successful completion.

\begin{align*}
&
\makeRule{
    \ruleFact{Check}{E, f(p), m}\\
    \ruleFact{!Pk}{M_i, pkM_i)}\\
}{
    \hspace{6ex}
}{
    \ruleFact{Check}{E, g(p), l(m)}\\
}%
\\ &
\makeRule{
    \ruleFact{Check}{E, p, \text{`'} }\\
    \ruleFact{!Pk}{A, pkA)}\\
}{
    \ruleFact{Complete}{E, f(p)}
}{
\hspace{2ex}-\hspace{2ex}\ 
}%
\end{align*}

As in similar rules, anonymous functions
$f, g, l$ are used to denote the changing
values of the payload and validation portions
of the message as it is decomposed between
verification steps. These functions are instantiated
for individual protocols.

The security definition 
for Verification-Dependent Path Integrity differs
only in the addition of an event marking that verification
was successful at the end of the protocol's execution. We
extend the final rule of the protocol with a
$\factFont{Complete}$ event fact, representing that
the verification process has successfully completed. This
event fact is parameterised by the path identifier, allowing
it to be readily associated with the corresponding
\factFont{StartBuild} fact. The corresponding security
claim differs only in that we include the existence of this
fact in the premise of the implication.

\section{Experiments}
\label{sec:case-studies}

To demonstrate the applicability of our framework, we 
consider multiple protocols from the literature. We
perform a security survey, building implementations
using the Tamarin~\cite{tamarin} prover
tool\footnote{Our Tamarin models, as well as
protocol diagrams showing our level of detail, are available from \url{https://github.com/path-integrity-analysis/path-integrity}}.

We split our analysis into three main families, as follows:\\
\textbf{Middlebox-Enabled TLS}, \textbf{Mixnets}
and \textbf{Payment Networks}. 
We consider
onion-style protocols (such as those demonstrated by
Chaum~\cite{chaummixnet}) as part of the Mixnet family --
although there can be some differences on the network
layer (such as by batching messages),
the message structure is often very similar.
Our reasons for
choosing these specific protocols are as follows:

\vspace{3.5ex}

\begin{itemize}
    \item \textbf{Middlebox-Enabled TLS}. mcTLS~\cite{mctls} uses session keys shared
    between multiple parties based on their
    permissions (read, write), 
    rather than their location in the path.
    mbTLS~\cite{mbtls} uses unique session 
    keys for each pair of adjacent agents.
    maTLS~\cite{matls} furthers this scheme with
    chained signatures (forming a \textit{modification
    log}).
    \item \textbf{Payment Networks}. The Lightning
    Network~\cite{lightning} uses per-hop payloads to conceal routing
    data, with a term $h(x)$ that is
    shared between agents. We break the protocol into two phases --
    the setup phase in which HTLCs are established, and the unlock
    phase in which funds are released by sharing the inverse hash $x$
    between participating agents. 
    \item \textbf{Mixnets}. The original models by
    Chaum~\cite{chaummixnet} use chained asymmetric
    encryption. The TOR~\cite{tor} ecosystem
    establishes connections using symmetric encryption
    (with public keys only for key-establishment) -
    paths are defined by a connection ID.
    HORNET~\cite{hornet} reduces the use of state
    compared to TOR, by including routing data as part
    of the message in place of a connection ID,
    using a construction based on Sphinx~\cite{sphinx}.

\end{itemize}

\vspace{3.5ex}


The results of our analysis can be found in
Table~\ref{fig:results-table}. 
We give results for Path Integrity (as per Definition~\ref{def:one-way-path-integrity}). For each protocol
we indicate if the security goal is met
($\alwaysSymb$) or violated ($\violatedSymb$).
An asterisk ($^*$) indicates the goal is dependent on
a Verification phase (as in Verification-Dependent
Path Integrity).



\begin{table}[tb]\centering
\caption{Security of protocols considered in our analysis}
\begin{tabular}{lc}
    \toprule
    \textbf{Protocol Name} & \textbf{Path Integrity} \\
    \midrule
    \textbf{Middlebox-Enabled TLS} &  \\
    \tableSpacer mcTLS~\cite{mctls} & \violatedSymb  \\
    \tableSpacer mbTLS~\cite{mbtls} & \violatedSymb  \\
    \tableSpacer maTLS~\cite{matls} & \verifSymb   \\
    \textbf{Payment Network} &  \\
    \tableSpacer Lightning (setup phase)~\cite{lightning} & \alwaysSymb  \\
    \tableSpacer Lightning (unlock phase)~\cite{lightning} & \violatedSymb  \\
    \textbf{Mixnet} &  \\
    \tableSpacer Chaum~\cite{chaummixnet} & \alwaysSymb  \\
    \tableSpacer TOR -- Establishment~\cite{tor} & \alwaysSymb  \\
    \tableSpacer TOR -- Data Exchange~\cite{tor} & \alwaysSymb  \\
    \tableSpacer HORNET~\cite{hornet} & \alwaysSymb \\
    \bottomrule
\end{tabular}
\label{fig:results-table}
\end{table}

\section{Conclusion}
\label{sec:conclusion}

In this paper we have considered multi-party protocols in which messages
are forwarded from one endpoint to another through a series of 
intermediate agents. We gave several formalisations
of path integrity inside a multiset rewriting model used by the automated
verification tool Tamarin.

We applied these definitions in a comprehensive security survey,
demonstrating a novel attack on mbTLS, and allowing for
simple, automated security proofs for payment networks on the Bitcoin
Lightning infrastructure and other scenarios. These attacks
demonstrate that path integrity is an important security property
that is not covered
by traditional authentication goals. Though the
impact of a message skipping attack can vary based on the domain,
the consequences can often be significant.

\mysubsect{Future Work}\\
There are several avenues for future work. 
Our approach focused on ensuring that agents behave in the correct
order as defined by a pre-set path. However, our analysis does not
consider traditional security goals, such as those
for secrecy or integrity.
Although there has been some work into approaches for the extension
of these security goals into multiparty
settings~\cite{casMultiparty},
this is often on a mutually pairwise basis.
Instead, considerations of the differing levels of participation of agents may lead to more precise
security definitions.

In this work we modelled the Lightning Payment protocol as two separate one-way
protocols. However, our model can be extended in order to enable analysis as a
singular protocol. This gives rise to a notion of \textit{symmetry} -- in the
case where the initial path of the message is not known or fixed, but we wish
to ensure that the return journey is identical to the forward journey. This may
be relevant in several fault-tolerant versions of payment network protocols, where
a payment is split into many small atomic transactions that are independently routed.


\bibliographystyle{ACM-Reference-Format}
\bibliography{main}




\end{document}